\documentclass[12]{article}
\usepackage{latexsym}
\usepackage{amsfonts}
\usepackage{amsmath}
\usepackage{amssymb}
\usepackage{amsthm}
\usepackage[latin1]{inputenc}

\makeatletter \@addtoreset{equation}{section}

\makeatletter
\newtheorem{theorem} {Theorem}
\newtheorem{lemma}{Lemma}
\newtheorem{definition} {Definition}

\newtheorem{remark}{Remark}
\newtheorem{example} {Example}

\input{tcilatex}

\begin{document}

\begin{center}
{\LARGE {\textbf{On the Klein-Gordon equation and hyperbolic pseudoanalytic function theory}} }
\end{center}

\vspace{1cm}

\begin{center}
{\large {V.V. Kravchenko$^{\dagger_1}$, D. Rochon$^{\dagger_2}$ and S.
Tremblay$^{\dagger_3}$} }
\end{center}

{\normalsize \bigskip }

\begin{center}
{\normalsize {$^{\dagger_1}$ CINVESTAV del IPN, Unidad Quer{\'e}taro,
Libramiento Norponiente No.~2000 C.P. 76230 Fracc. Real de Juriquilla, Quer{%
\'e}taro, Mexico\\[0pt]
E-mail: vkravchenko@qro.cinvestav.mx } }
\end{center}

{\normalsize \medskip }

\begin{center}
{\normalsize {$^{\dagger_2}$ D\'epartement de math\'ematiques et
d'informatique, Universit\'e du Qu\'ebec \`a Trois-Rivi\`eres, C.P. 500
Trois-Rivi\`eres, Qu\'ebec, Canada, G9A 5H7 \\[0pt]
E-mail: dominic.rochon@uqtr.ca} }
\end{center}

{\normalsize \medskip }

\begin{center}
{\normalsize {$^{\dagger_3}$ D\'epartement de math\'ematiques et
d'informatique, Universit\'e du Qu\'ebec \`a Trois-Rivi\`eres, C.P. 500
Trois-Rivi\`eres, Qu\'ebec, Canada, G9A 5H7 \\[0pt]
E-mail: sebastien.tremblay@uqtr.ca} }
\end{center}

{\normalsize \medskip }

\begin{abstract}
Elliptic pseudoanalytic function theory was considered independently by Bers and Vekua decades ago. In this
paper we develop a hyperbolic analogue of pseudoanalytic function theory using the algebra of hyperbolic
numbers. We consider the Klein-Gordon equation with a potential. With the aid of one particular solution we
factorize the Klein-Gordon operator in terms of two Vekua-type operators. We show that real parts of the
solutions of one of these Vekua-type operators are solutions of the considered Klein-Gordon equation. Using
hyperbolic pseudoanalytic function theory, we then obtain explicit construction of infinite systems of solutions
of the Klein-Gordon equation with potential. Finally, we give some examples of application of the proposed
procedure.
\end{abstract}

{\normalsize \vspace{4cm} \noindent \textbf{Keywords:} Pseudoanalytic Function Theory, Klein-Gordon Equation,
Hyperbolic Numbers.\newline }

{\normalsize \newpage }

\section{\protect\normalsize Introduction}

Pseudoanalytic function theory became an important part of classical complex
analysis after publication of the  monographs {\normalsize \cite{4, 17}
where the main basic results had been already established. In the recent
works \cite{8, 20} some new features and applications of this theory were
discovered, first of all an intimate relation between pseudoanalytic
functions and solutions of the stationary Schr{\"o}dinger equation as well as a
possibility to obtain explicitly complete systems of solutions for an ample
class of second-order elliptic partial differential equations. }

In the present work we develop a hyperbolic analogue of pseudoanalytic function theory which proves to be
extremely useful for studying hyperbolic partial differential equations. We show that solutions of the
Klein-Gordon equation with an arbitrary potential are closely related to certain hyperbolic pseudoanalytic
functions, the result of a factorization of the Klein-Gordon operator with the aid of two Vekua-type operators.
As one of the corollaries we obtain a method for explicit construction of infinite systems of solutions of the
considered Klein-Gordon equation. Our approach is based on the application of the algebra of hyperbolic numbers
\cite{14,16} instead of that of complex numbers and generalizes some earlier works dedicated to hyperbolic
analytic function theory {\normalsize \cite{32, 31, 18}. }

{\normalsize It should be mentioned that the elliptic and hyperbolic
pseudoanalytic function theories  naturally result to be quite different.
Nevertheless as we show in the present work there are many important common
features.}

\section{\protect\normalsize Hyperbolic numbers and analytic functions}

{\normalsize It has been proven (see, e.g.,  \cite{21}) that there exist
essentially three possible ways to generalize real numbers into real
algebras of dimension two. Indeed, each possible system can be reduced to
one of the following\newline
}

{\normalsize 1. numbers $a+b\mathrm{i}$ with $\mathrm{i}^2=-1$ (complex
numbers);\newline
}

{\normalsize 2. numbers $a+b\mathrm{j}$ with $\mathrm{j}^2=1$ (hyperbolic
numbers);\newline
}

{\normalsize 3. numbers $a+b\mathrm{k}$ with $\mathrm{k}^2=0$ (dual numbers).%
\newline
}

{\normalsize \noindent In this article the set of hyperbolic numbers, also
called duplex numbers (see, e.g., \cite{14,16}), will be denoted by }

{\normalsize
\begin{equation}
\mathbb{D}:=\big\{x+t\mathrm{j}\ :\ \mathrm{j}^{2}=1,\ x,t\in \mathbb{R}%
\big\}\cong \mathrm{Cl}_{\mathbb{R}}(0,1).  \label{defhyp}
\end{equation}%
It is easy to see that this algebra of hyperbolic numbers is commutative and
contains zero divisors. }

{\normalsize As in the case of complex numbers, we denote the real and
\textquotedblleft imaginary\textquotedblright\ parts of $z=x+t\mathrm{j}\in
\mathbb{D}$ by $x=\mathrm{Re}(z)$ and $t=\mathrm{Im}(z)$. Now, by defining
the conjugate as $\bar{z}:=x-t\mathrm{j}$ and the hyperbolic modulus as $%
|z|^{2}:=z\bar{z}=x^{2}-t^{2}$, we can verify that the inverse of $z$
whenever exists is given by
\begin{equation}
z^{-1}=\displaystyle\frac{\bar{z}}{|z|^{2}}.
\end{equation}%
From this, we find that the set $\mathcal{NC}$ of zero divisors of $\mathbb{D%
}$, called the \emph{null-cone}, is given by
\begin{equation*}
\mathcal{NC}=\big\{x+t\mathrm{j}\ :\ |x|=|t|\big\}.
\end{equation*}%
}

{\normalsize It is also possible to define differentiability of a function
at a point of $\mathbb{D}$ \cite{11, 12}: }

\begin{definition}
{\normalsize Let $U$ be an open set of $\mathbb{D}$ and $z_{0}\in U$. Then, $%
f:U\subseteq \mathbb{D}\longrightarrow \mathbb{D}$ is said to be $\mathbb{D}$%
-differentiable at $z_{0}$ with derivative equal to $f^{\prime }(z_{0})\in
\mathbb{D}$ if
\begin{equation}
\lim_{\overset{\scriptstyle z\rightarrow z_{0}}{\scriptscriptstyle(z-z_{0}%
\mbox{
}inv.)}}\frac{f(z)-f(z_{0})}{z-z_{0}}=f^{\prime }(z_{0}).
\end{equation}%
}
\end{definition}

Here $z$ tends to $z_{0}$ following the invertible trajectories.
{\normalsize We also say that the function $f$ is $\mathbb{D}$-holomorphic
on an open set $U$ if and only if $f$ is $\mathbb{D}$-differentiable at each
point of $U$. }

{\normalsize Any hyperbolic number can be seen as an element of $\mathbb{R}%
^{2}$, so a function $f(x+t\mathrm{j})=f_{1}(x,t)+f_{2}(x,t)\mathrm{j}$ can
be seen as a mapping $f(x,t)=\big(f_{1}(x,t),f_{2}(x,t)\big)$ of $\mathbb{R}%
^{2}$. }

\begin{theorem}
{\normalsize Let $U$ be an open set and $f:U\subseteq \mathbb{D}%
\longrightarrow \mathbb{D}$ such that $f\in {C}^{1}(U)$. Let also $f(x+t%
\mathrm{j})=f_{1}(x,t)+f_{2}(x,t)\mathrm{j}$. Then $f$ is $\mathbb{D}$%
-holomorphic on $U$ if and only if }

{\normalsize
\begin{equation}
\frac{\partial{f_1}}{\partial{x}} =\frac{\partial{f_2}}{\partial{t}}\ \ \ \
\ \mbox{ and }\ \ \ \ \ \frac{\partial{f_2}}{\partial{x}} =\frac{\partial{f_1%
}}{\partial{t}}.
\end{equation}
\newline
Moreover $f^\prime=\displaystyle
\frac{\partial{f_1}}{\partial{x}} +\displaystyle\frac{\partial{f_2}}{\partial%
{x}}\mathrm{j}$ and $f^\prime(z)$ is invertible if and only if $\det \mathcal{J%
}_{f}(z)\neq 0$, where $\mathcal{J}_f(z)$ is the Jacobian matrix of $f$ at $z$. \label{theobasic} }
\end{theorem}

{\normalsize \smallskip }

{\normalsize It is important to keep in mind that every hyperbolic number $%
x+t\mathrm{j}$ has the following unique idempotent representation
\begin{equation}
x+t\mathrm{j}=(x+t)\mathrm{e}_{1}+(x-t)\mathrm{e}_{2},  \label{idempotent}
\end{equation}%
where $\mathrm{e}_{1}=\displaystyle\frac{1+\mathrm{j}}{2}$ and $\mathrm{e}%
_{2}=\displaystyle\frac{1-\mathrm{j}}{2}$. This representation is very
useful because with its aid addition, multiplication and division can be
done term-by-term. }

{\normalsize The notion of holomorphicity can also be seen with this kind of
notation. For this we need to define the projections $P_{1},P_{2}:\mathbb{D}%
\longrightarrow \mathbb{R}$ as $P_{1}(z)=x+t$ and $P_{2}(z)=x-t$, where }$%
z=x+t\mathrm{j}$ {\normalsize as well as the following definition.}

\begin{definition}
{\normalsize We say that $X\subseteq\mathbb{D}$ is a $\mathbb{D}$-cartesian
set determined by $X_1$ and $X_2$ if
\begin{equation}
X=X_{1}\times_e X_{2}:=\big\{x+t\mathrm{j}\in\mathbb{D}:x+t\mathrm{j}=w_1%
\mathrm{e}_1+w_2\mathrm{e}_2, (w_1,w_2)\in X_1\times X_2\big\}.
\end{equation}
}
\end{definition}

{\normalsize It is easy to show that if $X_{1}$ and $X_{2}$ are open domains
of $\mathbb{R}$ then $X_{1}\times _{e}X_{2}$ is also an open domain of $%
\mathbb{D}$. Now, it is possible to formulate the following theorem.}

\begin{theorem}
{\normalsize If $f_{e_1}:X_1\longrightarrow \mathbb{R}$ and $%
f_{e_2}:X_2\longrightarrow \mathbb{R}$ are real differentiable functions on
the open domains $X_1$ and $X_2$ respectively, then the function $%
f:X_1\times_e X_2\longrightarrow \mathbb{D}$ defined as
\begin{equation}
f(x+t\mathrm{j})=f_{e_1}(x+t)\mathrm{e}_1+f_{e_2}(x-t)\mathrm{e}_2, \mbox{ }%
\forall\mbox{ }x+t\mathrm{j}\in X_1\times_e X_2
\end{equation}
is $\mathbb{D}$-holomorphic on the domain $X_1\times_e X_2$ and
\begin{equation}
f^\prime(x+t\mathrm{j})=f^\prime_{e_1}(x+t)\mathrm{e}_1+f^\prime_{e_2}(x-t)%
\mathrm{e}_2,\mbox{
}\forall\mbox{ }x+ t\mathrm{j} \in X_1\times_e X_2.
\end{equation}
\label{theo5} }
\end{theorem}

{\normalsize 
}

\section{\protect\normalsize Hyperbolic pseudoanalytic functions}

\subsection{\protect\normalsize Elementary hyperbolic derivative}

{\normalsize We will consider the variable $z=x+t\mathrm{j}$, where $x$ and $%
t$ are real variables and the corresponding formal differential operators
\begin{equation}
\partial _{z}=\frac{1}{2}\left( {\partial _{x}+\mathrm{j}\partial _{t}}%
\right) \mbox{ and }\partial _{\bar{z}}=\frac{1}{2}\left( {\partial _{x}-%
\mathrm{j}\partial _{t}}\right) .
\end{equation}%
Notation $f_{\bar{z}}$ or $f_{z}$ means the application of $\partial _{\bar{z%
}}$ or $\partial _{z}$ respectively to a hyperbolic function $f(z)=u(z)+v(z)%
\mathrm{j}$.
These hyperbolic operators act on sums, products, etc. just as an ordinary
derivative and we have the following result in the hyperbolic function
theory. We note that
\begin{equation*}
f_{z}=\frac{1}{2}\Big((u_{x}+v_{t})+(v_{x}+u_{t})\mathrm{j}\Big)\quad \mbox{and}\quad f_{\bar{z}}=\frac{1}{2}\Big((u_{x}-v_{t})+(v_{x}-u_{t})\mathrm{j}\Big).
\end{equation*}%

{\normalsize \noindent In view of these operators,
\begin{equation}
f_{z}(z)=0\ \ \ \Leftrightarrow\ \ \ (u_x+v_t)+(v_x+u_t)\mathrm{j}=0
\end{equation}
$\mbox{i.e. } u_x=-v_t,\ v_x=-u_t$ and
\begin{equation}
f_{\bar{z}}(z)=0\ \ \ \Leftrightarrow\ \ \ (u_x+v_t)+(v_x+u_t)\mathrm{j}=0
\end{equation}
$\mbox{i.e. } u_x=v_t,\ v_x=u_t.$ }

\begin{lemma}
{\normalsize Let $f(x+t\mathrm{j})=u(x,t)+v(x,t)\mathrm{j}$ be a hyperbolic
function where $u_x,u_t,v_x$ and $v_t$ exist, and are continuous in a
neighborhood of $z_0$. The derivative
\begin{equation}
f^{\prime}(z_0 ) =\lim_{\overset{\scriptstyle z \rightarrow z_{0}}{%
\scriptscriptstyle (z-z_{0}\mbox{
}inv.)}}\frac{f(z)-f(z_{0})}{z-z_{0}}
\end{equation}
exists, if and only if
\begin{equation}
f_{\bar{z}}(z_0)=0.  \label{CR2}
\end{equation}
Moreover, $f^{\prime}(z_0)=f_{z}(z_0)$ and $f^{\prime}(z_0)$ is invertible if and only if
$\det\mathcal{J}_{f}(z_0)\neq 0$. \label{basic} }
\end{lemma}

\subsection{\protect\normalsize Hyperbolic pseudoanalytic function theory}

{\normalsize Let $z=x+t\mathrm{j}$ where $x,t\in \mathbb{R}$. The theory is
based on assigning the part played by $1$ and $\mathrm{j}$ to two
essentially arbitrary hyperbolic functions $F$ and $G$. We assume that these
functions are defined and twice continuously differentiable in some open
domain $\Omega \subset \mathbb{D}$. We require that
\begin{equation}
\mbox{Im}\{\overline{F(z)}G(z)\}\neq 0.  \label{vec01}
\end{equation}%
Under this condition, $(F,G)$ will be called a generating pair in $\Omega $.
Notice that $\mbox{Im}\{\overline{F(z)}G(z)\}=\left\vert {%
\begin{array}{*{20}c}
{\mbox{Re}\{F(z)\}} & {\mbox{Re}\{G(z)\}}  \\
   {\mbox{Im}\{F(z)\}} & {\mbox{Im}\{G(z)\}}  \\
\end{array}}\right\vert .$ It follows, from Cramer's theorem, that for every
$z_{0}$ in $\Omega $ we can find unique constants $\lambda _{0},\mu _{0}\in
\mathbb{R}$ such that $w(z_{0})=\lambda _{0}F(z_{0})+\mu _{0}G(z_{0})$. More
generally we have the following result. }

\begin{theorem}
{\normalsize Let $(F,G)$ be generating pair in some open domain $\Omega $.
If $w(z):\Omega \subset \mathbb{D}\rightarrow \mathbb{D}$, then there exist
\textbf{unique} functions $\phi (z),\psi (z):\Omega \subset \mathbb{D}%
\rightarrow \mathbb{R}$ such that
\begin{equation}
w(z)=\phi (z)F(z)+\psi (z)G(z),\mbox{ }\forall z\in \Omega .
\end{equation}%
Moreover, we have the following explicit formulas for $\phi $ and $\psi $:
\begin{equation}
\phi (z)=\frac{\mathrm{Im}[\overline{w(z)}G(z)]}{\mathrm{Im}[\overline{F(z)}%
G(z)]}\mbox{, }\psi (z)=-\frac{\mathrm{Im}[\overline{w(z)}F(z)]}{\mathrm{Im}[%
\overline{F(z)}G(z)]}.
\end{equation}%
\label{explicit} }
\end{theorem}

{\normalsize \noindent \textbf{Proof.} Let $(F,G)$ be generating pair in
some open domain $\Omega $. Let $z_{0}\in \Omega $ with $w(z_{0})=x_{1}+t_{1}%
\mathrm{j}$, $F(z_{0})=x_{2}+t_{2}\mathrm{j}$ and $G(z_{0})=x_{3}+t_{3}%
\mathrm{j}$. In this case, $w(z_{0})=\phi (z_{0})F(z_{0})+\psi
(z_{0})G(z_{0})$ with $\phi (z_{0}),\psi (z_{0})\in \mathbb{R}$ if and only
if $x_{1}=\phi (z_{0})x_{2}+\psi (z_{0})x_{3}$ and $t_{1}=\phi
(z_{0})t_{2}+\psi (z_{0})t_{3}$. That is we obtain the system $AX=B$ where $%
A=\left(
\begin{smallmatrix}
x_{2} & x_{3} \\
t_{2} & t_{3}%
\end{smallmatrix}%
\right) $, $B=\left(
\begin{smallmatrix}
x_{1} \\
t_{1} \\
\end{smallmatrix}%
\right) $ and $X=\left(
\begin{smallmatrix}
\phi (z_{0}) \\
\psi (z_{0}) \\
\end{smallmatrix}%
\right) $ and the unique solution is $X=A^{-1}B$ where $A^{-1}=\frac{1}{{%
\det A}}\left(
\begin{smallmatrix}
t_{3} & -x_{3} \\
-t_{2} & x_{2}%
\end{smallmatrix}%
\right) $. Hence,
\begin{equation}
\begin{array}{rcl}
X & = & \displaystyle\frac{1}{{\mbox{Im}[\overline{F(z_{0})}G(z_{0})]}}%
\left( {\begin{array}{*{20}c} {t_3 } & { - x_3 } \\ { - t_2 } & {x_2 } \\
\end{array}}\right) \left( {\begin{array}{*{20}c} {x_1 } \\ {t_1 } \\
\end{array}}\right)  \\*[2ex]
& = & \displaystyle\frac{1}{{\mbox{Im}[\overline{F(z_{0})}G(z_{0})]}}\left( {%
\begin{array}{*{20}c} {\mbox{Im}[\overline{w(z_0)}G(z_0)]} \\
{-\mbox{Im}[\overline{w(z_0)}F(z_0)]} \\ \end{array}}\right) .%
\end{array}%
\end{equation}%
Then
\begin{equation}
\phi (z)=\frac{\mbox{Im}[\overline{w(z)}G(z)]}{\mbox{Im}[\overline{F(z)}G(z)]%
},\mbox{ }\psi (z)=-\frac{\mbox{Im}[\overline{w(z)}F(z)]}{\mbox{Im}[%
\overline{F(z)}G(z)]},\mbox{
}\forall z\in \Omega .\ \Box
\end{equation}%
}

{\normalsize Consequently, every hyperbolic function $w$ defined in some
subdomain of }$\Omega $ {\normalsize admits the unique representation $%
w=\phi F+\psi G$ where the functions $\phi $ and $\psi $ are real valued.
Thus, the pair $(F,G)$ generalizes the pair $(1,\mathrm{j})$ which
corresponds to hyperbolic analytic function theory. Sometimes it is
convenient to associate with the function $w$ the function $\omega =\phi +%
\mathrm{j}\psi $. The correspondence between $w$ and $\omega $ is
one-to-one. }

{\normalsize \smallskip \smallskip \hspace{0.5cm} We say that $w:\Omega
\subset \mathbb{D}\rightarrow \mathbb{D}$ possesses at $z_{0}$ the $(F,G)$%
-derivative $\dot{w}(z_{0})$ if the (finite) limit
\begin{equation}
\dot{w}(z_{0})=\lim_{\overset{\scriptstyle z\rightarrow z_{0}}{%
\scriptscriptstyle(z-z_{0}\mbox{
}inv.)}}\frac{{w(z)-\lambda _{0}F(z)-\mu _{0}G(z)}}{{z-z_{0}}}  \label{lim1}
\end{equation}%
exists. }

{\normalsize \hspace{0.5cm} The following expressions are called the
characteristic coefficients of the pair $(F,G)$:
\begin{equation}
\begin{array}{ll}
\label{coefficients} a_{(F,G)} =-\displaystyle \frac{\bar{F}G_{\bar{z}}-F_{%
\bar{z}}\bar{G}}{F\overline{G}-\overline{F}G}, & b_{(F,G)} =\displaystyle%
\frac{FG_{\bar{z}}-F_{\bar{z}}G}{F\overline{G}-\overline{F}G} \\*[2ex]
A_{(F,G)}=-\displaystyle\frac{\overline{F}G_{z}-F_{z}\overline{G}}{F%
\overline{G}-\overline{F}G}, & B_{(F,G)}=\displaystyle\frac{FG_{z}-F_{z}G}{F%
\overline{G}-\overline{F}G}.%
\end{array}%
\end{equation}
}

{\normalsize Set (for a fixed $z_{0}$)
\begin{equation}
W(z)=w(z)-\lambda _{0}F(z)-\mu _{0}G(z),
\end{equation}%
the constants $\lambda _{0},\mu _{0}\in \mathbb{R}$ being uniquely
determined by the condition
\begin{equation}
W(z_{0})=0.
\end{equation}%
Hence $W(z)$ has continuous partial derivatives if and only if $w(z)$ has.
Moreover, $\dot{w}(z_{0})$ exists if and only if $W^{\prime }(z_{0})$ does,
and if it does exist, then $\dot{w}(z_{0})=W^{\prime }(z_{0})$. Therefore,
by the Lemma \ref{basic}, if we suppose $w\in C^{1}(\Omega )$, the equation
\begin{equation}
W_{\bar{z}}(z_{0})=0  \label{cod}
\end{equation}%
is necessary and sufficient for the existence of (\ref{lim1}). Now,
\begin{equation}
W(z)=\frac{{\left\vert {\begin{array}{*{20}c} {w(z)} & {w(z_0 )} &
{\overline{w(z_0 )}} \\ {F(z)} & {F(z_0 )} & {\overline{F(z_0 )}} \\ {G(z)}
& {G(z_0 )} & {\overline{G(z_0 )}} \\ \end{array}}\right\vert }}{{\left\vert
{\begin{array}{*{20}c} {F(z_0 )} & {\overline{F(z_0 )}} \\ {G(z_0 )} &
{\overline{G(z_0 )}} \\ \end{array}}\right\vert }}
\end{equation}%
}

{\normalsize \noindent so that (\ref{cod}) may be written in the form
\begin{equation}
\left| {\begin{array}{*{20}c} {w_{\bar{z}}(z_0)} & {w(z_0 )} &
{\overline{w(z_0 )}} \\ {F_{\bar{z}}(z_0)} & {F(z_0 )} & {\overline{F(z_0
)}} \\ {G_{\bar{z}}(z_0)} & {G(z_0 )} & {\overline{G(z_0 )}} \\ \end{array} }
\right| = 0  \label{E1}
\end{equation}
and if (\ref{lim1}) exists, then
\begin{equation}
\dot w(z_0)= \frac{{\left| {\begin{array}{*{20}c} {w_z(z_0)} & {w(z_0 )} &
{\overline{w(z_0 )}} \\ {F_z(z_0)} & {F(z_0 )} & {\overline{F(z_0 )}} \\
{G_z(z_0)} & {G(z_0 )} & {\overline{G(z_0 )}} \\ \end{array} } \right|}} {{%
\left| {\begin{array}{*{20}c} {F(z_0 )} & {\overline{F(z_0 )}} \\ {G(z_0 )}
& {\overline{G(z_0 )}} \\ \end{array} } \right|}}.  \label{E2}
\end{equation}
Equations (\ref{E2}) and (\ref{E1}) can be rewritten in the form
\begin{equation}
\dot w=w_{z}-A_{(F,G)}w-B_{(F,G)}\overline{w}
\end{equation}
\begin{equation}
w_{\bar{z}}=a_{(F,G)}w+b_{(F,G)} \bar{w}.
\end{equation}
}

{\normalsize Thus we have proved the following result. }

\begin{theorem}
{\normalsize Let (F,G) be a generating pair in some open domain $\Omega $.
Every hyperbolic function $w\in C^{1}(\Omega )$ admits the unique
representation $w=\phi F+\psi G$ where $\phi ,\psi :\Omega \subset \mathbb{D}%
\rightarrow \mathbb{R}$. Moreover, the $(F,G)$-derivative $\dot{w}=%
\displaystyle\frac{d_{(F,G)}w}{dz}$ of $w(z)$ exists and has the form
\begin{equation}
\dot{w}=\phi _{z}F+\psi _{z}G=w_{z}-A_{(F,G)}w-B_{(F,G)}\overline{w}
\label{derivative}
\end{equation}%
if and only if
\begin{equation}
w_{\bar{z}}=a_{(F,G)}w+b_{(F,G)}\overline{w}.  \label{vekua}
\end{equation}%
}
\end{theorem}

{\normalsize \smallskip The equation (\ref{vekua}) can be rewritten in the
following form
\begin{equation}  \label{equivvekua}
\phi_{\bar{z}}F+\psi_{\bar{z}}G=0.
\end{equation}
Equation (\ref{vekua}) is called ``hyperbolic Vekua equation'' and any
continuously differentiable solutions of this equation are called
``hyperbolic $(F,G)$-pseudoanalytic functions''. }

{\normalsize If $w$ is hyperbolic $(F,G)$-pseudoanalytic, the associated function $%
\omega=\phi+\psi \mathrm{j}$ is called hyperbolic $(F,G)$-pseudoanalytic of second kind. }

\begin{remark}
{\normalsize The functions $F$ and $G$ are hyperbolic $(F,G)$-pseudoanalytic, and $\dot{F}\equiv \dot{G} \equiv
0$. }
\end{remark}

\begin{definition}
{\normalsize \label{DefSuccessor} Let $(F,G)$ and $(F_{1},G_{1})$ - be two
generating pairs in $\Omega$. $(F_{1},G_{1})$ is called \ successor of $%
(F,G) $ and $(F,G)$ is called predecessor of $(F_{1},G_{1})$ if%
\begin{equation*}
a_{(F_{1},G_{1})}=a_{(F,G)}\qquad\mbox{and}\qquad b_{(F_{1},G_{1})}=-B_{(F,G)}.
\end{equation*}
}
\end{definition}

{\normalsize The importance of this definition becomes obvious from the
following statement. }

\begin{theorem}
{\normalsize \label{ThBersDer} Let $w$ be a hyperbolic $(F,G)$-pseudoanalytic function and let $(F_{1},G_{1})$
be a successor of $(F,G)$.
If $\dot{w}=W\in C^{1}(\Omega )$ then $W$ is a hyperbolic $(F_{1},G_{1})$%
-pseudoanalytic function. }
\end{theorem}

{\normalsize \noindent \textbf{Proof.} The proof in the hyperbolic case is
identical to the elliptic case that we find in the book of Bers \cite{4}. }

\begin{definition}
{\normalsize \label{DefAdjoint}Let $(F,G)$ be a generating pair. Its adjoint
generating pair $(F,G)^{\ast}=(F^{\ast},G^{\ast})$ is defined by the formulas%
\begin{equation}  \label{F*G*}
F^{\ast}=-\frac{2\overline{F}}{F\overline{G}-\overline{F}G},\qquad G^{\ast }=%
\frac{2\overline{G}}{F\overline{G}-\overline{F}G}.
\end{equation}
}
\end{definition}

{\normalsize The $(F,G)$-integral is defined as follows
\begin{equation}
\int_{\Gamma }w\,\mathrm{d}_{(F,G)}z=F(z_{1})\func{Re}\int_{\Gamma }G^{\ast
}w\,\mathrm{d}z+G(z_{1})\func{Re}\int_{\Gamma }F^{\ast }w\,\mathrm{d}z
\label{integraldef}
\end{equation}%
where $\Gamma $ is a rectifiable curve leading from $z_{0}$ to $z_{1}$. }

{\normalsize If $w=\phi F+\psi G$ is a hyperbolic $(F,G)$-pseudoanalytic
function where $\phi $ and $\psi $ are real valued functions then
\begin{equation}
\int_{z_{0}}^{z}\dot{w}\,\mathrm{d}_{(F,G)}\zeta =w(z)-\phi (z_{0})F(z)-\psi
(z_{0})G(z).  \label{FGAnt}
\end{equation}%
This integral is path-independent and represents the $(F,G)$-antiderivative
of $\dot{w}$. The expression $\phi (z_{0})F(z)+\psi (z_{0})G(z)$ in (\ref%
{FGAnt}) can be seen as a \textquotedblleft \emph{pseudoanalytic constant}%
\textquotedblright\ of the generating pair $(F,G)$ in  $\Omega $. }

{\normalsize A continuous function $W(z)$ defined in a domain $\Omega $ will
be called $(F,G)$-integrable if for every closed curve $\Gamma $ situated in
a simply connected subdomain of $\Omega $ the following equality holds
\begin{equation}
\oint_{\Gamma }W\mathrm{d}_{(F,G)}z=0.  \label{FGintegrable}
\end{equation}%
}

{\normalsize 
}

\begin{theorem}
{\normalsize Let $W$ be a hyperbolic $(F,G)$-pseudoanalytic function. Then $W
$ is $(F,G)$-integrable. }
\end{theorem}

{\normalsize \noindent \textbf{Proof.} It will suffice to show that if $%
\Omega $ is a regular domain and $\Gamma $ lies within the domain of
definition of $W$, then
\begin{equation}
\int_{\Gamma }W\mathrm{d}_{(F,G)}z  \label{intWmaj}
\end{equation}%
is zero. }

{\normalsize From the definitions (\ref{coefficients}) and (\ref{F*G*}) we
find
\begin{equation}
\begin{array}{ll}
a_{(F^{\ast },G^{\ast })}=-a_{(F,G)}, & A_{(F^{\ast },G^{\ast })}=-A_{(F,G)},
\\*[2ex]
b_{(F^{\ast },G^{\ast })}=-\overline{B_{(F,G)}}, & B_{(F^{\ast },G^{\ast
})}=-\overline{b_{(F,G)}}.%
\end{array}%
\end{equation}%
Hence we obtain
\begin{equation}
F_{\bar{z}}^{\ast }=-aF^{\ast }-\overline{B}\,\overline{F^{\ast }},\ \ \ G_{%
\bar{z}}^{\ast }=-aG^{\ast }-\overline{B}\,\overline{G^{\ast }},
\end{equation}%
and by hypothesis
\begin{equation}
W_{\bar{z}}=aW-B\overline{W},
\end{equation}%
where $a,\ b,\ A$ and $B$ are the characteristics coefficients of $(F,G)$. }

{\normalsize Let us now use the definition (\ref{integraldef}) to evaluate (%
\ref{intWmaj}). By using the hyperbolic Green's theorem (see \cite{18}), we
obtain
\begin{equation*}
\begin{array}{rcl}
\displaystyle\int_{\Gamma }G^{\ast }W\mathrm{d}z & = & 2\mathrm{j}%
\displaystyle\int \int_{\Omega }\partial _{\bar{z}}(G^{\ast }W)\mathrm{d}x%
\mathrm{d}t \\*[2ex]
& = & 2\mathrm{j}\displaystyle\int \int_{\Omega }\Big(-aG^{\ast }W-\overline{%
B}\,\overline{G^{\ast }}W+G^{\ast }aW-G^{\ast }B\overline{W}\Big)\mathrm{d}x%
\mathrm{d}t \\*[2ex]
& = & -4\mathrm{j}\displaystyle\int \int_{\Omega }\mathrm{Re}\Big(G^{\ast }B%
\overline{W}\Big)\mathrm{d}x\mathrm{d}t%
\end{array}%
\end{equation*}%
which is a purely imaginary number. The same argument shows that $%
\int_{\Gamma }F^{\ast }W\mathrm{d}z$ is a pure imaginary number. Hence by
definition (\ref{integraldef}) we find that (\ref{intWmaj}) is zero . $\Box $
}

\subsection{{\protect\normalsize Generating sequences\label{SubsectGenSeq}}}

\begin{definition}
{\normalsize \label{DefSeq}A sequence of generating pairs $\big\{
(F_{m},G_{m})\big\}
$ with $m\in \mathbb{Z}$, is called a generating sequence if $%
(F_{m+1},G_{m+1})$ is a successor of $(F_{m},G_{m})$. If $%
(F_{0},G_{0})=(F,G) $, we say that $(F,G)$ is embedded in $\big\{%
(F_{m},G_{m})\big\}$. }
\end{definition}

{\normalsize 
}

\begin{definition}
{\normalsize A generating sequence $\big\{(F_{m},G_{m})\big\}$ is said to
have period $\mu>0$ if $(F_{m+\mu},G_{m+\mu})$ is equivalent to $%
(F_{m},G_{m})$ that is their characteristic coefficients coincide. }
\end{definition}

{\normalsize Let $w$ be a hyperbolic $(F,G)$-pseudoanalytic function. Using a generating sequence in which
$(F,G)$ is embedded we can define the higher
derivatives of $w$ by the recursion formula%
\begin{equation*}
w^{[0]}=w;\qquad w^{[m+1]}=\frac{\mathrm{d}_{(F_{m},G_{m})}w^{[m]}}{\mathrm{d%
}z},\quad m=1,2,\ldots
\end{equation*}
}

\begin{definition}
{\normalsize \label{DefFormalPower}The formal power $Z_{m}^{(0)}(a,z_{0};z)$
with center at $z_{0}\in \Omega $, coefficient $a$ and exponent $0$ is
defined as the linear combination of the generators $F_{m}$, $G_{m}$ with
real constant coefficients $\lambda $, $\mu $ chosen so that $\lambda
F_{m}(z_{0})+\mu G_{m}(z_{0})=a$. The formal powers with exponents $%
n=1,2,\ldots $ are defined by the recursion formula%
\begin{equation}
Z_{m}^{(n)}(a,z_{0};z)=n\int_{z_{0}}^{z}Z_{m+1}^{(n-1)}(a,z_{0};\zeta )%
\mathrm{d}_{(F_{m},G_{m})}\zeta .  \label{recformula}
\end{equation}%
}
\end{definition}

{\normalsize This definition implies the following properties. }

\begin{enumerate}
\item {\normalsize $Z_{m}^{(n)}(a,z_{0};z)$ is a $(F_{m},G_{m})$%
- hyperbolic pseudoanalytic function of $z$. }

\item {\normalsize If $a^{\prime }$ and $a^{\prime \prime }$ are real constants, then $Z_{m}^{(n)}(a^{\prime
}+\mathrm{j}a^{\prime \prime },z_{0};z)=a^{\prime }Z_{m}^{(n)}(1,z_{0};z)+a^{\prime \prime
}Z_{m}^{(n)}(\mathrm{j},z_{0};z).$ }

\item {\normalsize The formal powers satisfy the differential relations%
\begin{equation*}
\frac{\mathrm{d}_{(F_{m},G_{m})}Z_{m}^{(n)}(a,z_{0};z)}{dz}%
=nZ_{m+1}^{(n-1)}(a,z_{0};z).
\end{equation*}
}

\item {\normalsize The asymptotic formulas
\begin{equation*}
Z_{m}^{(n)}(a,z_{0};z)\sim a(z-z_{0})^{n},\quad z\rightarrow z_{0}
\end{equation*}
hold. }
\end{enumerate}

{\normalsize 
}

\section{\protect\normalsize Relationship between hyperbolic pseudoanalytic
functions and solutions of the Klein-Gordon equation}

\subsection{\protect\normalsize Factorization of the Klein-Gordon equation}

{\normalsize Consider the $(1+1)$-dimensional Klein-Gordon equation
\begin{equation}
\Big(\square -\nu (x,t)\Big)\varphi (x,t)=0  \label{Klein-Gordon}
\end{equation}%
in some domain $\Omega \subset \mathbb{R}^{2}$, where $\square :=%
\displaystyle\frac{\partial ^{2}}{\partial x^{2}}-\frac{\partial ^{2}}{%
\partial t^{2}}$, $\nu $ and $\varphi $ are real valued functions. We assume
that $\varphi $ is a twice continuously differentiable function.

As for the stationary two-dimensional Schr{\"o}dinger equation \cite{8} it is possible to factorize the Klein-Gordon
equation with potential. By $C$ we denote the the hyperbolic conjugation
operator.}
\begin{theorem}
{\normalsize \label{thmfacto} Let $f$ be a positive particular solution of (%
\ref{Klein-Gordon}) in $\Omega $. Then for any real valued function $\varphi
\in C^{2}(\Omega )$ the following equalities hold
\begin{equation}
\begin{array}{rcl}
(\square -\nu )\varphi  & = & 4\Big(\partial _{\bar{z}}+\displaystyle\frac{%
f_{z}}{f}C\Big)\Big(\partial _{z}-\displaystyle\frac{f_{z}}{f}C\Big)\varphi
\\*[2ex]
& = & 4\Big(\partial _{z}+\displaystyle\frac{f_{\bar{z}}}{f}C\Big)\Big(%
\partial _{\bar{z}}-\displaystyle\frac{f_{\bar{z}}}{f}C\Big)\varphi .%
\end{array}
\label{factorization}
\end{equation}%
}
\end{theorem}

{\normalsize \noindent \textbf{Proof.} Consider
\begin{equation}  \label{factorization1}
\begin{array}{rcl}
\Big(\partial_{\bar z}+\displaystyle \frac{f_z}{f}C\Big)\Big(\partial_{z}- %
\displaystyle \frac{f_z}{f}C\Big)\varphi & = & \partial_{\bar z}\partial_z
\varphi-\displaystyle \frac{|f_z|^2}{f^2}\varphi-\partial_{\bar z}\left(%
\displaystyle \frac{f_z}{f}\right)\varphi \\*[2ex]
& = & \displaystyle \frac{1}{4}(\square\, \varphi-\displaystyle \frac{%
\square f}{f}\varphi)=\frac{1}{4}(\square-\nu)\varphi.%
\end{array}%
\end{equation}
Thus we have the first equality in (\ref{factorization}). Now application of
$C$ to both sides of (\ref{factorization1}) gives us the second equality in (%
\ref{factorization}). $\Box$ }

{\normalsize Note that the operator $\partial _{z}-\frac{f_{z}}{f}I$, where $%
I$ is the identity operator, can be represented in the form
\begin{equation}
P=\partial _{z}-\frac{f_{z}}{f}I=f\partial _{z}f^{-1}I.
\end{equation}%
Let us introduce the notation $P:=f\partial _{z}f^{-1}I$. From Theorem \ref%
{thmfacto}, if $f$ is a positive solution of (\ref{Klein-Gordon}), the
operator $P$ transforms real valued solutions of (\ref{Klein-Gordon}) into
solutions of the following hyperbolic Vekua equation
\begin{equation}
\Big(\partial _{\bar{z}}+\frac{f_{z}}{f}C\Big)w=0.  \label{Vekua2}
\end{equation}%
}

{\normalsize The operator $\partial_z$ applied to a real valued function $%
\varphi$ can be regarded as a kind of gradient. If we have $\partial_z
\varphi=\Phi$ in a convex hyperbolic domain, where $\Phi=\Phi_1+\mathrm{j}%
\Phi_2$ is a given hyperbolic valued function such that its real part $%
\Phi_1 $ and imaginary part $\Phi_2$ satisfy
\begin{equation}  \label{condint1}
\partial_t \Phi_1-\partial_x\Phi_2=0,
\end{equation}
then we can construct $\varphi$ up to an arbitrary real constant $c$.
Indeed, we have
\begin{equation}  \label{intop1}
\varphi(x,t)=2\left(\int_{x_0}^x \Phi_1(\eta,t)\mathrm{d}\eta+\int_{t_0}^t
\Phi_2(x_0,\xi) \mathrm{d}\xi\right)+c
\end{equation}
where $(x_0,t_0)$ is an arbitrary fixed point in the domain of interest. We
will denote the integral operator in (\ref{intop1}) by $A$:
\begin{equation}  \label{intop2}
A[\Phi](x,t)=2\left(\int_{x_0}^x \Phi_1(\eta,t)\mathrm{d}\eta+\int_{t_0}^t
\Phi_2(x_0,\xi) \mathrm{d}\xi\right)+c
\end{equation}
Thus if $\Phi$ satisfies (\ref{condint1}), there exists a family of real
valued functions $\varphi$ such that $\partial_z\varphi=\Phi$, given by $%
\varphi=A[\Phi]$. }

{\normalsize In a similar way we define the operator $\bar{A}$ corresponding
to $\partial _{\bar{z}}$. The family of real-valued function $\varphi $ such
that $\partial _{\bar{z}}\varphi =\Phi $, where $\varphi =\overline{A}[\Phi ]
$, can be constructed as
\begin{equation}
\bar{A}[\Phi ](x,t)=2\left( \int_{x_{0}}^{x}\Phi _{1}(\eta ,t)\mathrm{d}\eta
-\int_{t_{0}}^{t}\Phi _{2}(x_{0},\xi )\mathrm{d}\xi \right) +c,
\label{intop3}
\end{equation}%
when
\begin{equation}
\partial _{t}\Phi _{1}+\partial _{x}\Phi _{2}=0.  \label{condint2}
\end{equation}%
Note that both definitions }$A$ and $\overline{A}$ are easily extended to
any simply connected domain.

{\normalsize Consider the operator $S=fAf^{-1}I$ applicable to any
hyperbolic valued function $w$ such that $\Phi=f^{-1}w$ satisfies condition (%
\ref{condint1}). Then it is clear that for such $w$ we have that $PSw=w$. }

\begin{theorem}
{\normalsize Let $f$ be a positive particular solution of (\ref{Klein-Gordon}%
) and $w$ be a solution of (\ref{Vekua2}). Then the real-valued function $%
g=Sw$ is a solution of (\ref{Klein-Gordon}). }
\end{theorem}

{\normalsize \noindent \textbf{Proof.} First, let us verify that the
function $\Phi=w/f$ satisfies condition (\ref{condint1}). Let $w=u+\mathrm{j}%
v$. We find
\begin{equation}  \label{expression}
\partial_t \Phi_1-\partial_x\Phi_2=f^{-1} \cdot \left((u_t-v_x)-\Big(\frac{%
f_t}{f}u-\frac{f_x}{f}v\Big)\right).
\end{equation}
The equation (\ref{Vekua2}) is equivalent to the system
\begin{equation}
u_x-v_t=-\displaystyle \frac{f_x}{f}u+\displaystyle \frac{f_t}{f}v,\ \ \
u_t-v_x=\displaystyle \frac{f_t}{f}u-\displaystyle \frac{f_x}{f}v.
\end{equation}
and we find that the expression (\ref{expression}) is zero. Hence, the
function $\Phi=w/f$ satisfies (\ref{condint1}) and we have $PSw=w$. }

{\normalsize Let $Q=(\partial_{\bar z}+\frac{f_z}{f}C)$ such that $QP=\frac{1%
}{4}(\square-\nu)$ from Theorem \ref{thmfacto}. We obtain
\begin{equation}
PSw=w \ \ \ \Rightarrow \ \ \ QPSw=Qw=0 \ \ \ \Rightarrow \ \ \ \frac{1}{4}%
(\square-\nu)Sw=0.\ \Box
\end{equation}
}

{\normalsize 
%
}

\subsection{\protect\normalsize The main hyperbolic Vekua equation}

{\normalsize The Vekua equation (\ref{Vekua2}) is closely related to another
Vekua equation given by }

{\normalsize
\begin{equation}
\Big(\partial _{\bar{z}}-\frac{f_{\bar{z}}}{f}C\Big)W=0.  \label{Vekua1}
\end{equation}%
Indeed, one can observe that the pair of functions
\begin{equation}
F=f\ \ \ \mbox{ and }\ \ \ G=\frac{\mathrm{j}}{f}  \label{genpair}
\end{equation}%
is a generating pair for (\ref{Vekua1}). The associated characteristic
coefficients are then given by
\begin{equation}
A_{(F,G)}=0,\ \ \ B_{(F,G)}=\displaystyle\frac{f_{z}}{f},\ \ \ a_{(F,G)}=0,\
\ \ b_{(F,G)}=\displaystyle\frac{f_{\bar{z}}}{f}  \label{carcoef}
\end{equation}%
and the $(F,G)$-derivative according to (\ref{derivative}) is defined as
follows}

{\normalsize
\begin{equation}
\ \dot{W}=W_z-\displaystyle \frac{f_z}{f}\overline{W}=\left(\partial_z-%
\displaystyle \frac{f_z}{f}C\right)W.
\end{equation}
}

{\normalsize From Definition \ref{DefSuccessor} and Theorem \ref{ThBersDer},
if we compare $B_{(F,G)}$ with the coefficient in (\ref{Vekua2}) we obtain
the following statement. }

\begin{theorem}
{\normalsize If $W\in C^{1}(\Omega)$ is a solution of (\ref{Vekua1}), then
its $(F,G)$-derivative $\dot W=w$ is a solution of (\ref{Vekua2}) on $\Omega$%
. }
\end{theorem}

{\normalsize Let us now consider the $(F,G)$-antiderivative. Taking into
account that $F^*=\mathrm{j}f$ and $G^*=1/f$, we find }

{\normalsize
\begin{equation}
\begin{array}{rcl}
\label{antiderivative}\displaystyle\int_{z_{0}}^{z}w(\zeta )\mathrm{d}%
_{(F,G)}\zeta  & = & \displaystyle f(z)\mathrm{Re}\int_{z_{0}}^{z}%
\displaystyle\frac{w(\zeta )}{f(\zeta )}\mathrm{d}\zeta -\displaystyle\frac{%
\mathrm{j}}{f(z)}\mathrm{Re}\int_{z_{0}}^{z}\mathrm{j}f(\zeta )w(\zeta )%
\mathrm{d}\zeta  \\*[2ex]
&  &  \\
& = & \displaystyle f(z)\mathrm{Re}\int_{z_{0}}^{z}\displaystyle\frac{%
w(\zeta )}{f(\zeta )}\mathrm{d}\zeta +\displaystyle\frac{\mathrm{j}}{f(z)}%
\mathrm{Im}\int_{z_{0}}^{z}f(\zeta )w(\zeta )\mathrm{d}\zeta
\end{array}%
\end{equation}%
and we obtain the following statement. }

\begin{theorem}
{\normalsize If $w$ is a solution of (\ref{Vekua2}), then the function
\begin{equation}
W(z)=\int_{z_0}^z w(\zeta)\mathrm{d}_{(F,G)}\zeta
\end{equation}
is a solution of (\ref{Vekua1}). }
\end{theorem}

\begin{lemma}
{\normalsize \label{bzreal} Let $b$ be a hyperbolic function such that $b_{z}
$ is a real-valued function, and let $W=u+\mathrm{j}v$ be a solution of the
equation
\begin{equation}
W_{\bar{z}}=b\overline{W}.
\end{equation}%
Then $u$ is a solution of the equation
\begin{equation}
\displaystyle\frac{1}{4}\square u=(b\bar{b}+b_{z})u
\end{equation}%
and $v$ is a solution of the equation
\begin{equation}
\displaystyle\frac{1}{4}\square v=(b\bar{b}-b_{z})v.
\end{equation}%
}
\end{lemma}

{\normalsize \noindent \textbf{Proof.} We observe that under the conjugation
the equation $W_{\bar z}=b\overline{W}$ is equivalent to $\partial_z(u-%
\mathrm{j}v)=\bar b(u+\mathrm{j}v)$. Then we obtain
\begin{equation}
\begin{array}{rcl}
\displaystyle \frac{1}{4}\square (u+\mathrm{j}v) & = & \partial_z\partial_{%
\bar z}(u+\mathrm{j}v) \\*[2ex]
& = & b_z (u-\mathrm{j}v)+b\partial_z(u-\mathrm{j}v) \\*[2ex]
& = & b_z(u-\mathrm{j}v)+b\bar b(u+\mathrm{j}v)%
\end{array}%
\end{equation}
and by considering the real and imaginary parts of this expression we
complete the proof. $\Box$ }

\begin{theorem}
{\normalsize \label{RealIma} Let $W$ be a solution of (\ref{Vekua1}). Then $%
u=\mathrm{Re}\ W$ is a solution of (\ref{Klein-Gordon}) and $v=\mathrm{Im}\ W
$ is a solution of the equation
\begin{equation}
\Big(\square -\eta \Big)v=0,\ \ \ \ \mbox{where }\ \ \ \eta =-\nu +8%
\displaystyle\frac{|f_{z}|^{2}}{f^{2}}.  \label{Klein-Gordon2}
\end{equation}%
}
\end{theorem}

{\normalsize \noindent \textbf{Proof.} Let us first show that for $b=%
\displaystyle \frac{f_{\bar z}}{f}$ then $b_z$ is a real-valued function:
\begin{equation}
b_z=\displaystyle\frac{(\partial_z f_{\bar z})f-f_{\bar z}f_z}{f^2}=\frac{1}{%
4}\displaystyle\frac{\square f}{f}- \displaystyle\frac{|f_{z}|^2}{f^2}=\frac{%
1}{4}\nu-\displaystyle\frac{|f_{z}|^2}{f^2}\in \mathbb{R}.
\end{equation}
We can easily calculate that $4(b\bar b +b_z)=\nu$ and $4(b\bar b -b_z)=\eta$
such that according to Lemma~\ref{bzreal} we find $(\square -\nu)u=0$ and $%
(\square -\eta)v=0$. $\Box$ }

\begin{remark}
{\normalsize If we consider the case $\nu=0$ in (\ref{Klein-Gordon}), then
we obtain the one-dimensional wave equation $\square \varphi=0$ with the
well known general solution $\varphi=F(x+t)+G(x-t)$, where $F$ and $G$ are
two arbitrary real-valued functions of one variable. In this case, the
potential $\eta$ is then given by $\eta=8\displaystyle \frac{%
F^{\prime}G^{\prime}}{(F+G)^2}$. }
\end{remark}

\begin{theorem}
{\normalsize Let $u$ be a solution of (\ref{Klein-Gordon}). Then the
function $v\in \ker (\square -\eta )$, such that $W=u+\mathrm{j}v$ is a
solution of (\ref{Vekua1}), is constructed according to the formula
\begin{equation}
v=-f^{-1}\overline{A}\big[\mathrm{j}f^{2}\partial _{\bar{z}}(f^{-1}u)\big].
\label{v}
\end{equation}%
It is unique up to an additive term $cf^{-1}$ where $c$ is an arbitrary real
constant. }

{\normalsize Let $v$ be a solution of (\ref{Klein-Gordon2}). Then the
function $u\in \ker (\square -\nu )$, such that $W=u+\mathrm{j}v$ is a
solution of (\ref{Vekua1}), can be constructed as
\begin{equation}
u=-f\overline{A}\big[\mathrm{j}f^{-2}\partial _{\bar{z}}(fv)\big],  \label{u}
\end{equation}%
up to an additive term $cf$. }
\end{theorem}

{\normalsize \noindent \textbf{Proof.} Consider $W=\phi f+\mathrm{j}\psi /f$
to be a solution of the Vekua equation (\ref{Vekua1}). Then this equation
can be rewritten in the form
\begin{equation}
\begin{array}{rcl}
\psi _{\bar{z}} & = & -\mathrm{j}f^{2}\phi _{\bar{z}} \\
& = & \displaystyle\frac{f^{2}}{2}(\phi _{t}-\mathrm{j}\phi _{x}).%
\end{array}%
\end{equation}%
Taking into account that $\phi =u/f$, $(\square -\nu )u=0$ and $(\square
-\nu )f=0$, we can verify that
\begin{equation}
\partial _{t}\left( \displaystyle\frac{f^{2}}{2}\phi _{t}\right) +\partial
_{x}\left( \displaystyle\frac{f^{2}}{2}\phi _{x}\right) =0,
\end{equation}%
such that we can use (\ref{intop3}) and $\psi $ is given by $\psi =-%
\overline{A}\big[\mathrm{j}f^{2}\phi _{\bar{z}}\big]$. Now, since $v=\mathrm{%
Im}\ W=\psi /f$ we find $v=-f^{-1}\overline{A}\big[\mathrm{j}f^{2}\partial _{%
\bar{z}}(f^{-1}u)\big]$. The function $v$ is a solution of (\ref%
{Klein-Gordon2}) due to Theorem~\ref{RealIma}. Note that as the operator $%
\overline{A}$ reconstructs the scalar function up to an arbitrary real
constant, the function $v$ in formula (\ref{v}) is uniquely determined up to
an additive term $cf^{-1}$ where $c$ is an arbitrary real constant. }

{\normalsize The equation (\ref{u}) is proved in a similar way. $\Box$ }

\begin{example}
{\normalsize \label{examplewithv} Let us illustrate the last theorem by a
simple example. Considering $f(x,t)=xt=\frac{1}{4}\big((x+t)^{2}-(x-t)^{2}%
\big)$ and $u(x,t)=1$ to be two particular solutions of the wave equation in
the subdomain $0<x<t<\infty $, then $v=-f^{-1}\overline{A}\big[\mathrm{j}%
f^{2}\partial _{\bar{z}}(f^{-1}u)\big]\in \ker \big(\square -\eta \big)$,
where
\begin{equation}
\eta (x,t)=8\displaystyle\frac{|f_{z}|^{2}}{f^{2}}=2\displaystyle\frac{%
t^{2}-x^{2}}{x^{2}t^{2}}.  \label{etapart}
\end{equation}%
Explicitly, the solution $v$ is given by
\begin{equation}
v(x,t)=\frac{x^{2}+t^{2}}{2xt}.
\end{equation}%
}
\end{example}

\subsection{\protect\normalsize Generating sequence of the main Vekua
equation}

{\normalsize The first step in the construction of a generating sequence for
the main Vekua equation (\ref{Vekua1}) is the construction of a generating
pair for the equation (\ref{Vekua2}) which, as was shown previously, is a
successor of the main Vekua equation. For this, one of the possibilities
consists in constructing another pair of solutions of (\ref{Vekua1}). Then
their $(F,G)$-derivatives will give us solutions of (\ref{Vekua2}). }

{\normalsize Consider the main Vekua equation (\ref{Vekua1}) which is
equivalent to the equation
\begin{equation}  \label{Vekuaequivalence}
\phi_{\bar z}F+\psi_{\bar z}G=0,
\end{equation}
where $W=\phi F+\psi G$, $F=f$ and $G=\mathrm{j}/f$. The equation (\ref%
{Vekuaequivalence}) can be rewritten explicitly as the following system of
partial differential equations
\begin{equation}  \label{pdesystem}
\begin{array}{rcl}
\phi_x f^2-\psi_t & = & 0, \\*[2ex]
\psi_x-\phi_t f^2 & = & 0.%
\end{array}%
\end{equation}
}

{\normalsize Let us suppose that $f$ and $\phi$ are functions of some real
variable $\rho=\rho(x,t)$, i.e. $f=f(\rho)$ and $\phi=\phi(\rho)$. The
system (\ref{pdesystem}) then becomes }

{\normalsize
\begin{equation}
\begin{array}{rcl}
\psi _{x} & = & \phi ^{\prime }\rho _{t}f^{2}, \\*[2ex]
\psi _{t} & = & \phi ^{\prime }\rho _{x}f^{2}.%
\end{array}
\label{pdesystem2}
\end{equation}%
The compatibility condition for this system implies
\begin{equation}
\partial _{x}\Big(\phi ^{\prime }\rho _{x}f^{2}\Big)-\partial _{t}\Big(\phi
^{\prime }\rho _{t}f^{2}\Big)=0,  \label{compatibility}
\end{equation}%
which is equivalent to the equation
\begin{equation}
\phi ^{\prime \prime }+\left( \displaystyle\frac{\square \rho }{4|\rho
_{z}|^{2}}+2\displaystyle\frac{f^{\prime }}{f}\right) \phi ^{\prime }=0,
\label{compatibility2}
\end{equation}%
for $|\rho _{z}|^{2}\neq 0$. We assume now that $\frac{\square \rho }{4|\rho
_{z}|^{2}}$ is a function of $\rho $, i.e.
\begin{equation}
s(\rho )=\displaystyle\frac{\square \rho }{4|\rho _{z}|^{2}}.  \label{smalls}
\end{equation}%
Hence, under this hypothesis, we can integrate (\ref{compatibility2}) and
obtain
\begin{equation}
\phi ^{\prime }(\rho )=\displaystyle\frac{\mathrm{e}^{-S(\rho )}}{f^{2}},
\label{phi'}
\end{equation}%
where $S(\rho )=\displaystyle\int_{\rho _{0}}^{\rho }s(\sigma )\,\mathrm{d}%
\sigma $. }

{\normalsize We can now integrate (\ref{phi'}) and (\ref{pdesystem2}) to
obtain a solution $W=\phi F+\psi G$ of (\ref{Vekua1}). However, since we are
interested to find a solution of (\ref{Vekua2}), i.e. the $(F,G)$-derivative
$\dot{W}$, we need $\phi _{z}$ and $\psi _{z}$ which are given explicitly by
\begin{equation}
\begin{array}{rcl}
\phi _{z} & = & \displaystyle\frac{\mathrm{e}^{-S}\rho _{z}}{f^{2}}, \\*[2ex]
\psi _{z} & = & \displaystyle\frac{\mathrm{j}}{2}\,\mathrm{e}^{-S}\rho _{z}.%
\end{array}
\label{phixpsiz}
\end{equation}%
Thus, a solution $w_{1}=\phi _{z}F+\psi _{z}G$ of (\ref{Vekua2}) is given by
\begin{equation}
w_{1}=\frac{3}{2}\,\mathrm{e}^{-S}\displaystyle\frac{\rho _{z}}{f}.
\label{w1}
\end{equation}%
}

{\normalsize In much the same way we can construct another solution of (\ref%
{Vekua2}) looking for $\psi=\psi(\rho)$. The system (\ref{pdesystem}) then
becomes }

{\normalsize
\begin{equation}  \label{pdesystem3}
\begin{array}{rcl}
\phi_x & = & \displaystyle\frac{\psi^{\prime}\rho_t}{f^2}, \\*[2ex]
\phi_t & = & \displaystyle\frac{\psi^{\prime}\rho_x}{f^2}.%
\end{array}%
\end{equation}
and $\psi^{\prime}(\rho)=f^2\mathrm{e}^{-S(\rho)}$. Calculating $\phi_z$ and
$\psi_z$ we find
\begin{equation}  \label{phixpsiz2}
\begin{array}{rcl}
\phi_z & = & \displaystyle \frac{\mathrm{j}}{2}\,\mathrm{e}^{-S}\rho_z, \\%
*[2ex]
\psi_z & = & f^2{\mathrm{e}}^{-S}\rho_z,%
\end{array}%
\end{equation}
which give us another solution $w_2$ of (\ref{Vekua2}):
\begin{equation}  \label{w2}
w_2=\frac{3}{2}\mathrm{j}\,\mathrm{e}^{-S}\rho_z f.
\end{equation}
}

{\normalsize Hence, for the function $\Phi=\mathrm{j}\,\mathrm{e}%
^{-S}\rho_z\neq 0$ we have found a generating pair for the Vekua equation (%
\ref{Vekua2}) given by (eliminating the constant $\frac{3}{2}$ in $w_1$ and $%
w_2$):
\begin{equation}  \label{F1G1}
\begin{array}{rcl}
\big(F_1,G_1\big) & = & \left(\mathrm{j}\,\mathrm{e}^{-S}\rho_z f,\ \mathrm{j%
}\,\mathrm{e}^{-S}\rho_z\displaystyle\frac{\mathrm{j}}{f}\right) \\*[2ex]
& = & \big(\Phi F,\Phi G\big).%
\end{array}%
\end{equation}
Indeed, we have
\begin{equation}  \label{ImF1G1}
\mathrm{Im}\big(\overline{F_1}G_1\big)=\mathrm{Im}\big(|\Phi|^2\overline{F}G%
\big)= -\mathrm{e}^{-S}|\rho_z|^2\neq 0.
\end{equation}
}

{\normalsize The following step is to construct the generating pair $%
(F_{2},G_{2})$. For this we should find two other solutions of (\ref{Vekua2}%
), equivalent to $\phi _{\bar{z}}F_{1}+\psi _{\bar{z}}G_{1}=0$. Then to
obtain $(F_{2},G_{2})$ we calculate the $(F_{1},G_{1})$-derivative of these
solutions. Using the same assumptions and the same method as in the previous
case, we obtain
\begin{equation}
\big(F_{2},G_{2}\big)=\big(\Phi ^{2}F,\Phi ^{2}G\big).  \label{F2G2}
\end{equation}%
The generalization of results (\ref{F1G1}) and (\ref{F2G2}) is given in the
next theorem which allows us to obtain a generating sequence wherein the
generating pair $(F,G)$ of (\ref{Vekua1}) is embedded. Let us note that
under the assumption (\ref{smalls}) the function $\Phi $ is a
\textquotedblleft hyperbolic analytic function\textquotedblright , i.e. $%
\Phi _{\bar{z}}=0$. Indeed, we have
\begin{equation*}
\begin{array}{rcl}
\Phi _{\bar{z}} & = & \mathrm{j}\Big((\partial _{\bar{z}}\mathrm{e}%
^{-S})\rho _{z}+\frac{1}{4}\mathrm{e}^{-S}\square \rho \Big) \\*[2ex]
& = & -\displaystyle\frac{1}{4}\mathrm{j}\,\mathrm{e}^{-S}\Big(4s|\rho
_{z}|^{2}-\square \rho \Big)=0.%
\end{array}%
\end{equation*}%
}

\begin{theorem}
{\normalsize Let $f$ be a nonvanishing solution of (\ref{Klein-Gordon}) such
that $f=f(\rho)$, $\rho=\rho(x,t)$, and $\frac{\square \rho}{4|\rho_z|^2}$
is a function of $\rho$ denoted by $s(\rho)$. Let also the function $\Phi$
such that $\Phi=\mathrm{j}\,\mathrm{e}^{-S(\rho)}\rho_z\neq 0$, where $%
S(\rho)=\int_{\rho_0}^{\rho}s(\sigma)\mathrm{d}\sigma$. Then the generating
pair $(F,G)$ with $F=f$ and $G=\mathrm{j}/f$ is embedded in the generating
sequence $(F_m,G_m)$ where $F_m=\Phi^m F$, $G_m=\Phi^m G$ and $m\in \mathbb{Z%
}$. }
\end{theorem}

{\normalsize \noindent \textbf{Proof.} First, let us show that $(F_m,G_m)$
is a generating pair in $\mathbb{Z}$. Indeed, we find
\begin{equation*}
\mathrm{Im}\big(\overline{F_m} G_m\big)=\mathrm{Im}\big(|\Phi|^{2m}\overline{%
F}G\big)= (-1)^{m}\mathrm{e}^{-2mS}|\rho_z|^{2m}\neq 0.
\end{equation*}
}

{\normalsize To complete the proof, we need to show that $\big\{(F_{m},G_{m})%
\big\}$ forms a generating sequence, i.e. $(F_{m},G_{m})$ is a successor of $%
(F_{m-1},G_{m-1})$:
\begin{equation}
a_{(F_{m},G_{m})}=a_{(F_{m-1},G_{m-1})}\ \ \ \mbox{and}\ \ \
b_{(F_{m},G_{m})}=-B_{(F_{m-1},G_{m-1})}.  \label{proofgeneseq}
\end{equation}%
The coefficients $a_{(F_{m},G_{m})}$, $b_{(F_{m},G_{m})}$ and $%
B_{(F_{m},G_{m})}$ can be calculated in terms of $a_{(F,G)}$, $b_{(F,G)}$
and $B_{(F,G)}$, respectively, by taking into account that $\Phi _{\bar{z}}=0
$. We obtain
\begin{equation}
\begin{array}{l}
a_{(F_{m},G_{m})}=|\Phi |^{2m}a_{(F,G)}=0,\ \ \ \ b_{(F_{m},G_{m})}=%
\displaystyle\left( \frac{\Phi }{\overline{\Phi }}\right) ^{m}b_{(F,G)}, \\%
*[2ex]
B_{(F_{m},G_{m})}=\displaystyle\left( \frac{\Phi }{\overline{\Phi }}\right)
^{m}B_{(F,G)}.%
\end{array}
\label{abB}
\end{equation}%
Therefore, the first equality in (\ref{proofgeneseq}) is verified. Taking
into account (\ref{carcoef}) and (\ref{abB}), the second equality in (\ref%
{proofgeneseq}) is reduced to
\begin{equation}
\overline{\Phi }f_{z}+\Phi f_{\bar{z}}=0\ \ \Leftrightarrow \ \ f^{\prime }(%
\overline{\Phi }\rho _{z}+\Phi \rho _{z})=0.  \label{eqfinproof}
\end{equation}%
Since $\Phi =\mathrm{j}\,\mathrm{e}^{-S(\rho )}\rho _{z}$ it is easy to
observe that (\ref{eqfinproof}) is valid. $\Box $ }

{\normalsize 
}

This last theorem allow us to calculate the generating sequence $(F_m,G_m)$ for a large class of  potentials
$\nu(x,t)$ in the Klein-Gordon equation (\ref{Klein-Gordon}). The importance of this result appears in the
following theorem.

\begin{theorem}
\label{solutionthm} {\normalsize Let $f$ be a particular solutions of (\ref{Klein-Gordon}) and let $(F,G)$ be
the generating pair in some open domain $\Omega$ with $F=f$ and $G=\mathrm{j}/f$. Then
$$
\mathrm{Re}\ Z^{(n)}(a,z_0;z),\ \ \ \ n=0,1,2,\ldots
$$
are solutions of the Klein-Gordon equation (\ref{Klein-Gordon}). }
\end{theorem}
\noindent \textbf{Proof.} From property 1 of the definition \ref{DefFormalPower} we see that $Z^{(n)}(a,z_0;z)$
is a hyperbolic $(F,G)$-pseudoanalytic function. Hence $Z^{(n)}(a,z_0;z)$ satisfies (\ref{Vekua1}) and the real
parts are solutions of (\ref{Klein-Gordon}) from Theorem \ref{RealIma}. $\Box$

\begin{example}
{\normalsize As an example of this theorem, we consider the Klein-Gordon
equation (\ref{Klein-Gordon}) with the potential $\nu (x,t)=t^{2}-x^{2}$ in
the \textquotedblleft time-like\textquotedblright\ subdomain $0<x<t<\infty $%
. A particular solution of this equation is given by $f(\rho )=\mathrm{e}%
^{\rho ^{2}}$, where we have defined $\rho (x,t)=\sqrt{xt}$. In this case
the function $\frac{\square \rho }{4|\rho _{z}|^{2}}$ is a function of $\rho
$ given by $s(\rho )=-1/\rho $, with $S(\rho )=-\ln \rho $ and $\Phi =\frac{z%
}{4}\neq 0$. Let us construct the first formal powers $Z^{(n)}(1,4\mathrm{j}%
;z)$ and $Z^{(n)}(\mathrm{j},4\mathrm{j};z)$. By the definition \ref%
{DefFormalPower} we have
\begin{equation*}
\begin{array}{rclrcl}
Z^{(0)}(1,4\mathrm{j};4\mathrm{j}) & = & 1, & Z^{(0)}(\mathrm{j},4\mathrm{j}%
;4\mathrm{j}) & = & \mathrm{j}, \\
& = & \lambda _{1}F(4\mathrm{j})+\mu _{1}G(4\mathrm{j}), &  & = & \lambda
_{2}F(4\mathrm{j})+\mu _{2}G(4\mathrm{j}), \\
& = & \lambda _{1}+\mathrm{j}\mu _{1}, &  & = & \lambda _{2}+\mathrm{j}\mu
_{2},%
\end{array}%
\end{equation*}%
such that $\lambda _{1}=\mu _{2}=1$ and $\mu _{1}=\lambda _{2}=0$. Hence, we
obtain
\begin{equation*}
\begin{array}{rclrcl}
Z^{(0)}(1,4\mathrm{j};z) & = & \lambda _{1}F(z)+\mu _{1}G(z), & Z^{(0)}(%
\mathrm{j},4\mathrm{j};z) & = & \lambda _{2}F(z)+\mu _{2}G(z) \\
& = & \mathrm{e}^{xt}, &  & = & \mathrm{j}\mathrm{e}^{-xt}.%
\end{array}%
\end{equation*}%
Now, from the formula (\ref{recformula}), if we want to construct $%
Z^{(1)}(1,4\mathrm{j};z)$ and $Z^{(1)}(\mathrm{j},4\mathrm{j};z)$ we need to
calculate first $Z_{1}^{(0)}(1,4\mathrm{j};z)$ and $Z_{1}^{(0)}(\mathrm{j},4%
\mathrm{j};z)$. First note that the generating pair $(F_{1},G_{1})$ is given
by
\begin{equation*}
F_{1}=\frac{1}{4}z\mathrm{e}^{xt}\ \ \ \ \mbox{and}\ \ \ \ G_{1}=\frac{%
\mathrm{j}}{4}z\mathrm{e}^{-xt}.
\end{equation*}%
Hence, from definition \ref{DefFormalPower} we obtain
\begin{equation*}
\begin{array}{rclrcl}
Z_{1}^{(0)}(1,4\mathrm{j};4\mathrm{j}) & = & 1, & Z_{1}^{(0)}(\mathrm{j},4%
\mathrm{j};4\mathrm{j}) & = & \mathrm{j}, \\
& = & \lambda _{3}F_{1}(4\mathrm{j})+\mu _{3}G_{1}(4\mathrm{j}), &  & = &
\lambda _{4}F_{1}(4\mathrm{j})+\mu _{4}G_{1}(4\mathrm{j}), \\
& = & \lambda _{3}\mathrm{j}+\mu _{3}, &  & = & \lambda _{4}\mathrm{j}+\mu
_{4},%
\end{array}%
\end{equation*}%
which implies that $\mu _{3}=\lambda _{4}=1$ and $\lambda _{3}=\mu _{4}=0$
and
\begin{equation*}
\begin{array}{rclrcl}
Z_{1}^{(0)}(1,4\mathrm{j};z) & = & \lambda _{3}F_{1}(z)+\mu _{3}G_{1}(z), &
Z_{1}^{(0)}(\mathrm{j},4\mathrm{j};z) & = & \lambda _{4}F_{1}(z)+\mu
_{4}G_{1}(z) \\
& = & \displaystyle\frac{\mathrm{j}}{4}z\mathrm{e}^{-xt}, &  & = & %
\displaystyle\frac{\mathrm{1}}{4}z\mathrm{e}^{xt}.%
\end{array}%
\end{equation*}%
>From definition (\ref{recformula}), we obtain
\begin{equation*}
Z^{(1)}(a,4\mathrm{j};z)=\int_{0}^{z}Z_{1}^{(0)}(a,4\mathrm{j};\zeta )%
\mathrm{d}_{(F,G)}\zeta
\end{equation*}%
and (\ref{F*G*}) gives $F^{\ast }=\mathrm{j}f$ and $G^{\ast }=1/f$. Now using (\ref{integraldef}) we find
\begin{equation*}
\begin{array}{rcl}
Z^{(1)}(1,4\mathrm{j};z) & = & \displaystyle\frac{1}{4}\left( \mathrm{e}^{xt}%
\mathrm{Re}\displaystyle\int_{0}^{z}\mathrm{j}\mathrm{e}^{-2x^{\prime
}t^{\prime }}\zeta d\zeta +\mathrm{j}\mathrm{e}^{-xt}\mathrm{Re}\displaystyle%
\int_{0}^{z}\zeta\mathrm{d}\zeta \right) , \\*[2ex] Z^{(1)}(\mathrm{j},4\mathrm{j};z) & = &
\displaystyle\frac{1}{4}\left(
\mathrm{e}^{xt}\mathrm{Re}\displaystyle\int_{0}^{z}\zeta d\zeta +\mathrm{j}%
\mathrm{e}^{-xt}\mathrm{Re}\displaystyle\int_{0}^{z}%
\mathrm{j}\mathrm{e}^{2x't'}\zeta \mathrm{d}\zeta \right) ,%
\end{array}%
\end{equation*}%
where $\zeta =x^{\prime }+\mathrm{j}t^{\prime }$. Evaluating these
integrals, we obtain
\begin{equation*}
\mathrm{Re}\displaystyle\int_{0}^{z}\zeta \mathrm{d}\zeta =\mathrm{Re}%
\displaystyle\int_{0}^{1}\epsilon (x+\mathrm{j}t)(x+\mathrm{j}t)\mathrm{d}%
\epsilon =\displaystyle\frac{x^{2}+t^{2}}{2}
\end{equation*}%
and
\begin{equation*}
\mathrm{Re}\displaystyle\int_{0}^{z}\mathrm{j}\mathrm{e}^{\pm 2x^{\prime
}t^{\prime }}\zeta \mathrm{d}\zeta =\mathrm{Re}\displaystyle\int_{0}^{1}%
\mathrm{j}(x+\mathrm{j}t)^{2}\epsilon \mathrm{e}^{\pm 2\epsilon ^{2}xt}%
\mathrm{d}\epsilon =\mathrm{e}^{\pm xt}\sinh (xt),
\end{equation*}%
such that
\begin{equation*}
\begin{array}{rcl}
Z^{(1)}(1,4\mathrm{j};z) & = & \displaystyle\frac{1}{4}\left( \sinh (xt)+%
\mathrm{j}\frac{x^{2}+t^{2}}{2}\mathrm{e}^{-xt}\right) , \\*[2ex]
Z^{(1)}(\mathrm{j},4\mathrm{j};z) & = & \displaystyle\frac{1}{4}\left( \frac{%
x^{2}+t^{2}}{2}\mathrm{e}^{xt}+\mathrm{j}\sinh (xt)\right) .%
\end{array}%
\end{equation*}%
}

{\normalsize Now, from the formula (\ref{recformula}), if we want to find $%
Z^{(2)}(1,4\mathrm{j};z)$ and $Z^{(2)}(\mathrm{j},4\mathrm{j};z)$ we need to
calculate first $Z^{(1)}_1(1,4\mathrm{j};z)$ and $Z^{(1)}_1(\mathrm{j},4%
\mathrm{j};z)$; those are themselves obtained from $Z^{(0)}_2(1,4\mathrm{j}%
;z)$ and $Z^{(0)}_2(\mathrm{j},4\mathrm{j};z)$. }

{\normalsize The generating pair $(F_{2},G_{2})$ is given by
\begin{equation*}
F_{2}=\left(\frac{z}{4}\right)^2\mathrm{e}^{xt}\ \ \ \ \mbox{and}\ \ \ \
G_{2}=\mathrm{j}\left(\frac{z}{4}\right)^2\mathrm{e}^{-xt},
\end{equation*}%
which allows us to calculate $Z_{2}^{(0)}(1,4\mathrm{j};z)$ and $Z_{2}^{(0)}(%
\mathrm{j},4\mathrm{j};z)$. We find }

{\normalsize
\begin{equation*}
\begin{array}{ll}
Z^{(0)}_2(1,4\mathrm{j};z)=\displaystyle \left(\frac{z}{4}\right)^2 \mathrm{e}^{xt}, &
Z^{(0)}_2(\mathrm{j},4\mathrm{j};z)=\displaystyle \mathrm{j}\left(\frac{z}{4}\right)^2
\mathrm{e}^{-xt}.%
\end{array}%
\end{equation*}
To obtain $Z^{(1)}_1(1,4\mathrm{j};z)$ and $Z^{(1)}_1(\mathrm{j},4\mathrm{j}%
;z)$, we need the adjoint generating pair of $(F_1,G_1)$, i.e. }

{\normalsize
\begin{equation*}
F_1^*=4\mathrm{j}\displaystyle \frac{\mathrm{e}^{xt}}{z},\ \ \ \ G_1^*=4%
\displaystyle \frac{\mathrm{e}^{-xt}}{z}.
\end{equation*}
Using (\ref{recformula}), we find }

{\normalsize
\begin{equation*}
\begin{array}{rcl}
Z_{1}^{(1)}(1,4\mathrm{j};z) & = & \frac{1}{16}(x+\mathrm{j}t)\left( \frac{%
x^{2}+t^{2}}{2}\mathrm{e}^{xt}+\mathrm{j}\sinh (xt)\right)  \\*[2ex]
Z_{1}^{(1)}(\mathrm{j},4\mathrm{j};z) & = & \frac{1}{16}(x+\mathrm{j}%
t)\left( \sinh (xt)+\mathrm{j}\frac{x^{2}+t^{2}}{2}\mathrm{e}^{-xt}\right) .%
\end{array}%
\end{equation*}%
Finally, by considering again (\ref{recformula}), we obtain }

{\normalsize
\begin{equation*}
\begin{array}{rcl}
Z^{(2)}(1,4\mathrm{j};z) & = & \frac{\mathrm{e}^{xt}}{64}\Big[%
(x^{2}+t^{2})^{2}+4xt+2\big(\cosh (2xt)-\sinh (2xt)-1\big)\Big] \\*[2ex]
&  & +\frac{\mathrm{j}}{64}\frac{x^{2}+t^{2}}{xt}\Big[\mathrm{e}^{xt}\big(%
4xt\sinh (2xt)-1\big) \\*[2ex]
&  & +2\mathrm{e}^{-xt}\cosh (xt)\big(\cosh (xt)+\sinh (xt)\big)-1\Big], \\%
*[2ex]
Z^{(2)}(\mathrm{j},4\mathrm{j};z) & = & -\frac{1}{64}\frac{x^{2}+t^{2}}{xt}%
\Big[\mathrm{e}^{xt}\big(\sinh (2xt)-\cosh (2xt)\big)-4xt\sinh (xt)+\mathrm{e%
}^{-xt}\Big] \\*[2ex] &  & -\frac{\mathrm{j}}{64}\mathrm{e}^{-xt}\Big[(x^{2}+t^{2})^{2}+4\cosh
(xt)\big(\sinh (xt)+\cosh (xt)\big)-4\big(xt+1\big)\Big].%
\end{array}%
\end{equation*}%
}

Using Theorem \ref{solutionthm}, we can now verify that the real parts of $Z^{(n)}(1,4\mathrm{j};z)$ and
$Z^{(n)}(\mathrm{j},4\mathrm{j};z)$ with $n=0,1,2,\ldots$ are solutions of the Klein-Gordon equation with
potential $\nu(x,t)=t^2-x^2$.
\end{example}

{\normalsize
\begin{example}
We are now considering the Klein-Gordon equation (\ref{Klein-Gordon}) with potential
\begin{equation}
\nu(x,t)=\displaystyle \frac{1}{4}\left(\displaystyle \frac{1}{(t+1)^2}-\displaystyle \frac{1}{(x+1)^2}\right)
\label{potexample3}
\end{equation}
on the time-like subdomain $0<x<t<\infty$. A particular solution of this equation is given by
$f(x,t)=\sqrt{(x+1)(t+1)}$. We denote $\rho=(x+1)(t+1)$. In this case it is easy to see that the function
$\frac{\square \rho }{4|\rho _{z}|^{2}}$ is zero, therefore a function of $\rho$. We obtain
$\Phi=\frac{z}{2}+\mathrm{e}_1$, where $\mathrm{e}_1$ is the idempotent constant of (\ref{idempotent}). Let us
calculate the first formal powers $Z^{(n)}(1,t_0\mathrm{j};z)$ and $Z^{(n)}(\mathrm{j},t_0\mathrm{j};z)$, where
$t_0>0$. By definition~7 we find
$$
Z^{(0)}(1,t_0\mathrm{j};z)=\alpha^{-1}\sqrt{(x+1)(t+1)},\ \ \
Z^{(0)}(\mathrm{j},t_0\mathrm{j};z)=\frac{\mathrm{j}\alpha}{\sqrt{(x+1)(t+1)}}.
$$
where $\alpha=\sqrt{t_0+1}$. From (\ref{recformula}), in order to construct $Z^{(1)}(1,t_0\mathrm{j};z)$ and
$Z^{(1)}(\mathrm{j},t_0\mathrm{j};z)$ we first need $Z^{(0)}_1(1,t_0\mathrm{j};z)$ and
$Z^{(0)}_1(\mathrm{j},t_0\mathrm{j};z)$. These functions are calculating from the generating pair $(F_1,G_1)$
given by
$$
F_1(z)=\left(\frac{z}{2}+\mathrm{e}_1\right)\sqrt{(x+1)(t+1)}, \ \ \
G_1(z)=\left(\frac{\mathrm{j}z}{2}+\mathrm{e}_1\right)\frac{1}{\sqrt{(x+1)(t+1)}}.
$$
Using this generating pair and by definition~7 we obtain
$$
\begin{array}{rcl}
Z^{(0)}_1(1,t_0\mathrm{j};z)&=&-\displaystyle \frac{z+2\mathrm{e}_1}{\alpha t_0(t_0+2)}\sqrt{(x+1)(t+1)}+
\mathrm{j}\displaystyle \frac{\alpha (t_0+1)(z+2\mathrm{e}_1)}{t_0(t_0+2)}\frac{1}{\sqrt{(x+1)(t+1)}}\\*[2ex]
Z^{(0)}_1(\mathrm{j},t_0\mathrm{j};z)&=&\displaystyle \frac{(t_0+1)(z+2\mathrm{e}_1)}{\alpha t_0(t_0+2)}\sqrt{(x+1)(t+1)}
  -\mathrm{j}\displaystyle \frac{\alpha(z+2\mathrm{e}_1)}{t_0(t_0+2)}\frac{1}{\sqrt{(x+1)(t+1)}}.
\end{array}
$$
We note that $F^*=-\mathrm{j}\sqrt{(x+1)(t+1)}$ and $G^*=1/\sqrt{(x+1)(t+1)}$ such that we are now able to calculate
$Z^{(1)}(1,t_0\mathrm{j};z)$ and $Z^{(1)}(\mathrm{j},t_0\mathrm{j};z)$. We find
$$
\begin{array}{l}
Z^{(1)}(1,t_0\mathrm{j};z)=\sqrt{(x+1)(t+1)}\left[\frac{-1}{\alpha
t_0(t_0+2)}\left(\frac{x^2+t^2}{2}+x+t\right)+\frac{\alpha
(t_0+1)}{t_0(t_0+2)}\ln\left[(x+1)(t+1)\right]\right]\\*[2ex]
+\frac{\mathrm{j}}{\sqrt{(x+1)(t+1)}}\left[\frac{1}{2\alpha
t_0(t_0+2)}\Big(2(x+t)(xt+1)+(x^2t^2+4xt+x^2+t^2)\Big)\right.\\*[2ex] \left. -
\frac{\alpha(t_0+1)}{t_0(t_0+2)}\left(\frac{x^2+t^2}{2}+x+t\right)\right]
\end{array}
$$
and
$$
\begin{array}{l}
Z^{(1)}(\mathrm{j},t_0\mathrm{j};z)=\sqrt{(x+1)(t+1)}\left[\frac{t_0+1}{\alpha
t_0(t_0+2)}\left(\frac{x^2+t^2}{2}+x+t\right)-\frac{\alpha}{t_0(t_0+2)}\ln\left[(x+1)(t+1)\right]\right]\\*[2ex]
+\frac{\mathrm{j}}{\sqrt{(x+1)(t+1)}}\left[\frac{-(t_0+1)}{2\alpha
t_0(t_0+2)}\Big(2(x+t)(xt+1)+(x^2t^2+4xt+x^2+t^2)\Big)\right.\\*[2ex] \left. +
\frac{\alpha}{t_0(t_0+2)}\left(\frac{x^2+t^2}{2}+x+t\right)\right].
\end{array}
$$
Again here, we can verify that the real parts of $Z^{(1)}(1,t_0\mathrm{j};z)$ and
$Z^{(1)}(\mathrm{j},t_0\mathrm{j};z)$ are solutions of the Klein-Gordon equation with potential
(\ref{potexample3}).
\end{example}
}

\section{Conlusions}

We proved that the Klein-Gordon equation can be reduced to a hyperbolic
Vekua equation of the form (\ref{Vekua1}) and as a consequence under quite
general conditions an infinite system of solutions of the Klein-Gordon
equation can be constructed explicitly as a real part of the corresponding
set of formal powers. Meanwhile in the elliptic theory this result gave us
\cite{20} a complete system of solutions (of a corresponding Schr{\"o}dinger
equation) it is an open question what part of the kernel of the Klein-Gordon
operator is determined by the obtained solutions.

One of the main results of elliptic pseudoanalytic function theory is the so-called similarity principle
\cite{4,1,3}. It is interesting and important to find what is a corresponding fact in the hyperbolic case.

The reduction of the Klein-Gordon equation with an arbitrary potential to a Vekua-type hyperbolic first order
equation gives the possibility to apply concepts and ideas from pseudoanalytic function theory to linear
second-order wave equations. Besides some first applications presented in this work, questions related to
initial and boundary value problems, existence and construction of special classes of solutions, large-time
behaviour of solutions (closely related to a similarity principle) and others may receive a new development
effort.

\section*{Acknowledgments}

D.R. and S.T. thank the CINVESTAV del IPN in Quer{\'e}taro for hospitality. V.K. acknowledges the support of CONACYT
of Mexico via the research project 50424. The research of D.R. is partly supported by grants from CRSNG of
Canada and FQRNT of Qu{\'e}bec. The research of S.T. is partly supported by grant from CRSNG of Canada.

\end{document}